\documentclass[a4paper,10pt]{aa}

\usepackage{graphicx, color}
\usepackage{txfonts}
\usepackage{natbib} 
\bibpunct{(}{)}{;}{a}{}{,}
\begin{document}
\title{Structure of sunspot light bridges in the chromosphere and transition region}

\author{R. Rezaei$^{1,2}$}

\authorrunning{R.~ Rezaei 2017}  
   \institute{Instituto de Astrof\'isica de Canarias (IAC), V\'ia Lact\'ea, 38200 La Laguna (Tenerife), Spain
     \and Departamento de Astrof\'isica, Universidad de La Laguna, 38205 La Laguna (Tenerife), Spain }

\keywords{Sun: sunspots -- Sun: chromosphere -- Sun: transition region}

\abstract
{Light bridges (LBs) are elongated structures with enhanced intensity embedded in sunspot umbra and pores.} 
{We studied the properties of a sample of $60$ LBs  observed with the Interface Region Imaging Spectrograph (IRIS)\rm.} 
{Using IRIS near- and far-ultraviolet spectra, we measured the line intensity, width, and Doppler shift; followed traces of LBs in the 
chromosphere and transition region (TR); and compared LB parameters with  umbra and quiet Sun.} 
{There is a systematic emission enhancement in LBs compared to nearby umbra 
from the photosphere up to the TR.
Light bridges are systematically displaced toward the solar limb at higher layers: 
the amount of the displacement at one solar radius 
compares well with the typical height  of the chromosphere and TR. 
The intensity of the LB sample compared to the umbra sample 
peaks at the middle/upper chromosphere where they are almost permanently bright. 
Spectral lines emerging from the LBs are broader than the nearby umbra.  
The systematic redshift of the \ion{Si}{IV} line in the LB sample is reduced compared to the quiet Sun sample.
We found a significant correlation between the line width of ions arising at temperatures from $3\times10^4$ to $1.5\times10^5$K 
as there is also a strong spatial correlation among the line and continuum intensities. 
In addition, the intensity$-$line width relation holds for all spectral lines in this study. 
The correlations indicate that the cool and hot plasma in LBs are coupled.} 
{Light bridges comprise multi-temperature and multi-disciplinary structures extending up to the TR.
Diverse heating sources supply the energy and momentum to different layers, resulting in distinct dynamics in 
the photosphere, chromosphere, and TR.}
\maketitle

\section{Introduction}
Sunspots are manifestations of high concentrations of magnetic flux in the solar photosphere.  
They live up to several months, orders of magnitude longer than a convection timescale 
($\approx$\,$10$--$20$\,min). Sunspots are structured in a dark umbra surrounded by a penumbra. 
The magnetic field lines are close to vertical in an umbra and close to horizontal in a penumbra.
Larger umbra are typically darker and show higher magnetic field strength \citep{bray_loughhead_64,mcintosh_1981}.
Two magnetic configurations are discussed for sunspots: a monolithic flux tube and a cluster of flux
tubes \citep{parker_1979_spot_formation, spruit_1981a}. 
The formation of sunspots have not often been observed due to the short timescale of
the formation process compared to the lifetime of sunspots \citep{rolf_etal_2010a, shimizu_etal_2012}. 
Sunspot properties are summarized in several review papers
\citep{solanki03r,moradi_etal_2010, rempel_schlichenmaier_2011, borrero_ichimoto_2011}.

Sunspot umbra have an average intensity of $\approx\!60\,\%$ in the near-infrared and  
$\approx\!1\,\%$ in the near-ultraviolet (UV) continuum, in units of the average quiet Sun intensity. 
Sunspot umbra usually consist of a dark umbral core, light bridge(s), and umbral dots.
Light bridges (LBs) are elongated regions with enhanced intensities compared to nearby umbra.  
They either split an umbra or are embedded in the umbra. 
\citet{muller_1979} categorized LBs based on their photospheric continuum intensities: 
photospheric, penumbral, or umbral. 
From previous observations, it is known that LBs  generally have a weaker magnetic field strength 
and a more horizontal field orientation than nearby
umbra \citep{lites_etal_1991, wiehr_degenhardt_1993, leka1997, jurcak_etal_2006,felipe_etal_2016}. 

Light bridges show transient brightenings in higher atmospheric layers
during the formation and decay of sunspots \citep{zwaan92, shimizu_etal_09, reza_etal_2012a}. 
They have either a segmented or a filamentary morphology \citep{lites_etal_04}. 
They are  observed to form through the continuation of penumbra filaments into the umbra \citep{katsukawa_etal_07}.
Light bridges can also form as trapped granules by the coalescence of pores during the formation and growth of sunspots. 
These LBs can reappear during the decay of a sunspot \citep{vazquez_1973, garcia_87}. 
Signs of the overturning convection are observed in LBs, leading observers to conclude that the convection 
is the main energy transport mechanism in LBs \citep{rimmele_1997, rimmele_08, roupe_luis_etal_2010, lagg_etal_2014}. 
Thermodynamic properties of the LBs in pores are reported by \citet{hirzberger_etal_2002}, 
\citet{giordano_etal_2008}, and \citet{sobotka_etal_2013}. 
In his numerical simulation, \cite{rempel_2011} found a LB associated with deep structures
reaching down to the bottom of the simulation domain (> 10\,Mm). 
In some high-resolution observations, a dark lane was observed in LBs \citep{sobotka_etal93, berger_berdyugina_2003}, 
which was reproduced in a simulation of an emerging active region by \citet{cheung_etal10}.

Anomalous, unidirectional, or bidirectional\,(shear) flows are observed in LBs 
\citep{berger_berdyugina_2003,schleicher_etal_2003,rimmele_08,louis_etal_2009,roupe_luis_etal_2010,rolf_etal_2012,bello_gonzalez_etal_2012}.
The chromospheric activity of LBs are observed in the line-core images of the \ion{Ca}{ii}\,H and H$\alpha$ lines 
\citep{shimizu_etal_09, roubustini_etal_2016}. In \ion{Ca}{ii} filtergrams, LBs show small-scale jets
of several arcseconds moving away from the LB \citep{louis_etal_2014, beck_etal_2015}. 
H$\alpha$ surges have also been observed associated with LBs \citep{canfield_etal_1996, asai_etal_2001, tziotziou_etal_2005}.

Partially ionized plasma in the chromosphere and transition region (TR) is usually investigated through the 
UV spectrum of diverse ionized atoms forming at a temperature range of $4\!\le\!\rm{\log\!T[K]}\le\!6$.  
The NASA Interface Region Imaging Spectrograph (IRIS) is a small telescope launched on June 27, 2013, that 
observes the Sun  in the UV part of the spectrum \cite[IRIS, ][]{iris_2013}.
IRIS records spectral lines forming at a range of temperature
and heights that spans the photosphere up to the middle of the TR.
In the near-UV, IRIS observes a spectral range around the two strong \ion{Mg}{ii}
lines at $280$\,nm and their nearby continuum, while
the far-UV spectra includes several strong emission lines originating in the solar TR ($133$-$141$\,nm). 
Compared to the Solar Ultraviolet Measurements of Emitted 
Radiation \citep[SUMER, ][]{summer}, IRIS records a restricted spectral
range with a higher spatial and spectral resolution and with four UV slitjaw channels.

The temperature rise above the photosphere is accompanied by an increase in the nonthermal (turbulence) velocity.  
Profiles of the chromospheric and TR lines have been analyzed to measure the temperature and nonthermal
velocities as well as the electron density \citep{tandberg_1960, moe_nicolas_1977, lemaire_2007}. 
The TR lines usually show some degree of asymmetry as well as
a large line width far in excess of their thermal broadening. 
The deviation of the wing intensities of TR lines from a single Gaussian cannot be explained in terms of 
the damping since the electron densities are not high enough. 
Multi-component models are required to fit the average quiet Sun line profiles,
indicating an inhomogeneous structure of the solar TR \citep{peter_2000, peter_2010}. 
\citet{mariska_1992} reviews the solar TR while 
\citet{aschwanden_2004} summarizes the relevant techniques and physical processes in the TR and corona.

In this paper, we study the mere existence and physical properties of light bridges  
in the solar chromosphere and transition region and try to establish a 
connection between the intensity enhancements in different layers.
Currently a cusp-shape magnetic configuration is supported by photospheric observations of LBs 
\citep{leka1997, jurcak_etal_2006,lagg_etal_2014,felipe_etal_2016}. 
In the cusp model, the umbral field lines wrap around the nonmagnetic or weak field 
light bridge and merge in the upper photosphere \citep[][their Fig.~7]{jurcak_etal_2006}. 
These magnetic field lines cannot be differentiated from umbral fields in the higher layers.  
Therefore, we do not expect to find trace of LBs in the chromosphere and TR (similar to umbral dots). 
In other words, the existence of significant LB signals in the upper atmosphere indicates
a systematic effect, and challenges the cusp model. 
Compiling a large sample of LBs, we collected evidence of the presence of a 
LB signal in the upper atmosphere. 
To this end we observed $40$ sunspots with IRIS including $60$ LBs. 
Using IRIS data which is free from seeing and differential refraction effects,
we studied LB images in the photosphere, chromosphere, and TR. 
The paper has the following structure. 
We review the observed datasets and briefly discuss the analysis methods in Sect.~\ref{sec:obs}.
The sample of $42$\,k LB pixels gathered from the observed LBs 
allows us to evaluate the line and continuum intensities, the line width and the Doppler shifts. 
The distribution of these parameters, the correlations among them, and a comparison of the
LB sample with the umbra and quiet Sun samples are addressed in Sect.~\ref{sec:result}. 
Discussion and conclusions are presented in Sects.~\ref{sec:discussion} and \ref{sec:conclusion}, respectively. 
A follow-up paper will address IRIS slitjaws and magnetic properties of the light bridges.

\begin{table}
\begin{center}
\caption{IRIS observations. $x$ and $y$ stand for the solar heliographic coordinate.
Time denotes the start time of the scan in UT while $t$ shows the integration time per slit position. 
In total, there are 60 LBs in our sample. The number of LBs in each observation is given in the last column. 
Horizontal lines separate different active regions.}\vspace{-0.1cm}
\label{tab:sample}
\begin{tabular}{llllllll}\hline
entry & date & time  & NOAA & $x\,[\arcsec]$ & $y\,[\arcsec]$ & $t$\,[s] & $\#$\\\hline\hline
1 & 20130814   &  18:45  & 11817  & +216   & -435  & 5  & 3\\\hline
2 & 20131012   &  20:01  & 11861  & +45    & -251  & 9  & 2\\\hline
3 & 20131207   &  03:09  & 11916  & +207   & -229  & 5  & 2\\\hline
4 & 20131220   &  16:27  & 11930  & -217   & -167  & 5  & 2\\\hline
5 & 20140131   &  20:21  & 11967  & -398   & -154  & 32 & 1\\\hline
6 & 20140503   &  15:52  & 12049  & +85    & -85   & 16 & 3\\\hline
7 & 20140511   &  18:32  & 12059  & -32    & +105  & 16 & 1\\\hline
8 & 20140517   &  11:50  & 12063  & -142   & +231  & 9  & 1\\\hline
9 & 20140703   &  08:00  & 12104  & -322   & -224  & 5  & 2\\\hline   
10 & 20140724a &  15:00  & 12121  & -683   &  +62  & 32  & 2\\
11 & 20140724b &  18:35  &  12121 & -704  &  +51  & 32  & 2\\
12 & 20140725  &  09:30 &  12121 &  -575  &  +67  & 32  & 1\\
13 & 20140727  &  14:02 &  12121 &  -119  &  +40  & 32  & 1\\
14 & 20140728  &  16:02 &  12121 &  +111  &  +42  & 32  & 1\\
15 & 20140729  &  15:00 &  12121 &  +328  &  +50  & 32  & 1\\
16 & 20140730  &  15:04 &  12121 &  +584  &  -225 & 32  & 1\\\hline     
17 & 20140731  &  15:03 &  12127 &  -412  &  -211 & 32  & 1\\
18 & 20140801  &  17:20 &  12127 &  -220  &  -230 & 32  & 1\\\hline
19 & 20140820   & 05:40 &  12146 &  -569  &  +83  & 5   & 1\\\hline
20 & 20141022   & 18:33 &  12192 &  -246  &  -306 & 32  & 2\\
21 & 20141024   & 18:52 &  12192 &  +180  &  -328 & 16  & 2\\
22 & 20141025a  & 07:00 &  12192 &  +262  &  -303 & 16  & 3\\
23 & 20141025b  & 21:00 &  12192 &  +400  &  -315 & 16  & 2\\
24 & 20141027   & 18:57 &  12192 &  +724  &  -282 & 16  & 1\\\hline
25 & 20150213   & 06:11 &  12282 &  -220  &   291 & 6   & 1\\\hline    
26 & 20150327a  & 13:59 &  12305 &  +119  &  -50  & 32  & 1\\
27 & 20150327b  & 17:33 &  12305 &  +121  &  -30  & 32  & 1\\
28 & 20150328a  & 13:59 &  12305 &  +328  &  -66  & 32  & 1\\
29 & 20150328b  & 16:42 &  12305 &  +336  &  -45  & 32  & 2\\
30 & 20150329   & 13:59 &  12305 &  +529  &  -74  & 32  & 1\\
31 & 20150330   & 13:59 &  12305 &  +690  &  -92  & 32  & 1\\\hline     
32 & 20150420   & 07:59 &  12325 &  +26   &  +155 & 5   & 1\\\hline     
33 & 20151216   & 20:26 &  12470 &  -470  &  +218 & 10  & 1\\
34 & 20151220   & 10:33 &  12470 &  +296  &  +236 & 10  & 1\\\hline     
35 & 20151224   & 23:27 &  12473 &  -641  &  -335 & 10  & 3\\\hline 
36 & 20160108   & 15:58 &  12480 &  -744  &  +84  & 10  & 2\\\hline 
37 & 20160414   & 04:27 &  12529 &  +52   &  +269 & 9   & 1\\
38 & 20160415   & 11:12 &  12529 &  +329  &  +264 & 9   & 1\\\hline     
39 & 20160901   & 14:09 &  12585 &  -811  &  +77  & 17  & 2\\
40 & 20160904   & 06:27 &  12585 &  -385  &  +35  & 17  & 1\\\hline 
\end{tabular}
\end{center}
\end{table}
\begin{figure}[t]
\resizebox{\hsize}{!}{\includegraphics*{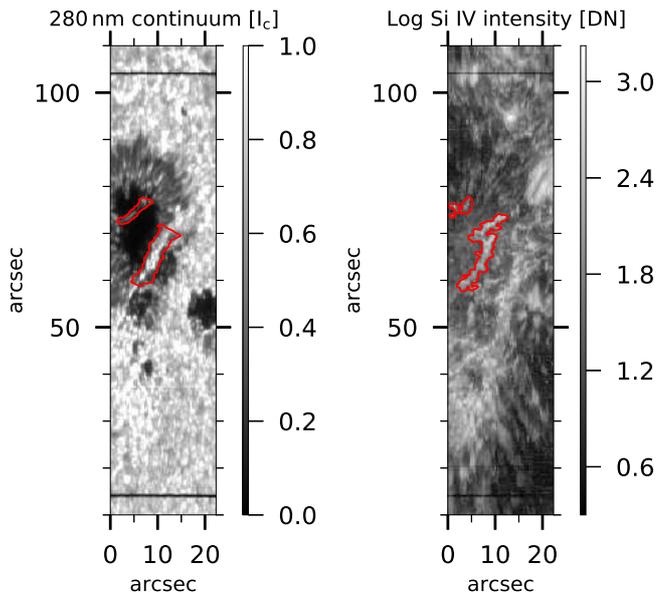}}
\caption{ IRIS maps of observation \#$10$ in Table~\ref{tab:sample}.
  The left panel displays the photospheric continuum map normalized to
  the average quiet Sun intensity. The right panel shows the logarithm of the intensity 
  of the \ion{Si}{iv}\,$140.3$\,nm  line measured with a
  Gaussian fit (see Sect.~\ref{sect:fuv_spectrum} for more details). 
  The $280$\,nm and silicon masks of the two LBs are marked on the corresponding panels.
  Both maps were clipped for a better visibility. The widths of the LBs are comparable in the photosphere and the TR. 
  We note the relative shift of the LBs in the two maps. The solar limb is toward the left.}
\label{fig:mask}
\end{figure}
\section{Observations and data analysis}\label{sec:obs}
Table\ref{tab:sample} lists properties of IRIS sunspot scans.
Our sample of $40$ sunspots  observed between $2013$ and $2016$ comprises $60$ LBs. 
The integration time varies from $5$\,s to $32$\,s, leading to a
substantial difference in the signal-to-noise ratios among different datasets.
All intensities discussed through the paper (except the one in Fig.~\ref{fig:mask}) 
are normalized by the integration time to make a comparison possible. 
Each map was analyzed in the full slit length; we have the two hair lines
in the analyzed parameters and use them to align the data in different spectral channels (Fig.~\ref{fig:mask}). 
In the following, we only discuss spectral lines that are present in all the scans in our sample. 

All observed spectra except for the $280$\,nm band  were subject to a median filtering and
a convolution with a small Gaussian kernel to reduce the effect of noise, which  
is particularly important when dealing with a low-amplitude signal like that in the \ion{O}{i}\,$135.56$\,nm line.
Cosmic rays (CRs) are corrected on two-dimensional rasters after masking spectral lines. Whenever 
a line was hit by a CR, the amplitude was compared to similar lines to discriminate between a strong signal and a CR.
The method was conservative, so the low-amplitude CRs pass through this filter, 
and we did not modify spectral lines when it was not necessary.

We created three masks for each LB using an intensity threshold: the continuum at the $280$\,nm, the \ion{Mg}{ii}\,k line core,
and the \ion{Si}{iv} core intensities. The sum of the magnesium and silicon masks constructed an overall mask for each 
LB. The intensity threshold for the $280$\,nm mask was set to $0.1$\,I$_{\mathrm{c}}$,
where  I$_{\mathrm{c}}$ stands for the average quiet Sun intensity.  
For the silicon mask, the threshold was set to $3$\,DN (Fig~\ref{fig:mask}).  
For the \ion{O}{i} and \ion{C}{ii} lines, the overall mask was used. 
We gathered analyzed data from all LBs inside the masks, resulting in $42,436$ entries for each line parameter. 
Sample IRIS maps with two LBs are shown in Fig.~\ref{fig:mask}. As seen in this example, 
the width of the LBs in the photosphere and the TR are similar (LBs do not expand exponentially with height). 
In the following, we discuss statistics of the line intensity, the line width, and the Doppler shift  
in this sample that includes all individual pixels in our LBs. 
To facilitate a comparison of the LB sample with quiet Sun and umbra, we created two extra samples. 
These two statistical ensembles consist of a large fraction of quiet Sun/umbra regions in our IRIS observations. 
The umbra sample does not include penumbral grains or LBs but consists of bright and dark umbra (umbral flashes). 
There are more entries in the quiet Sun and umbra samples than in the LB sample ($2,502,255$ and $194,783$, respectively).
The pixel size of IRIS spectral rasters in the far-UV (near-UV) is $0.33$ ($0.4$)\arcsec, while 
the slit width is $0.33$\arcsec \citep{iris_2013}.

\subsection{Near-ultraviolet spectrum}
The near-Ultraviolet (NUV) spectrograph records a spectral range
around the \ion{Mg}{ii}\,h and k lines and two quasicontinuum windows in the $280$\,nm range. 
The spectral dispersion in the NUV channel was either $2.544$ or $5.088$\,pm 
corresponding to a velocity dispersion of $2.7$ or $5.5$\,km\,s$^{-1}$ per pixel, respectively. 
The velocity modulations due to the orbital motion of the satellite was corrected for 
using a strong photospheric line in the $280$\,nm channel (\ion{Fe}{i}\,$283.2435$\,nm). 
To this end, we calculated the average line-core position along each slit. 
The average spectral position as a function of time was then fitted with
a polynomial to make sure that the residual systematic variation
is below the expected uncertainty of the velocity calibration. 
When a systematic trend was clear along the slit
(particularly for observations with the full slit length of $175$\arcsec\,\,far from the disk center),
a linear fit was used to correct for this spectral gradient. 
For a velocity calibration in the photosphere,
we assumed the average quiet Sun velocity to be zero, leading to a systematic error
of less than $1$\,km\,s$^{-1}$, which is small compared to the velocity dispersion. 
\begin{figure}
\resizebox{\hsize}{!}{ \includegraphics*{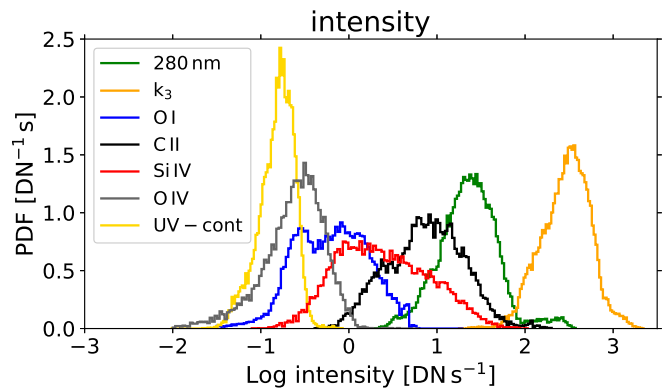}}
\caption{Histograms of the line and continuum intensities in the LB sample.
Intensities are normalized to the integration time.
The core intensity of the \ion{Mg}{ii}\,k line is denoted k$_{\rm{3}}$.
The LB sample includes $42,436$ entries from $60$ LBs. We used $100$ bins in all histograms.}
\label{fig:hist_int}
\end{figure}
\begin{figure}
\resizebox{\hsize}{!}{ \includegraphics*{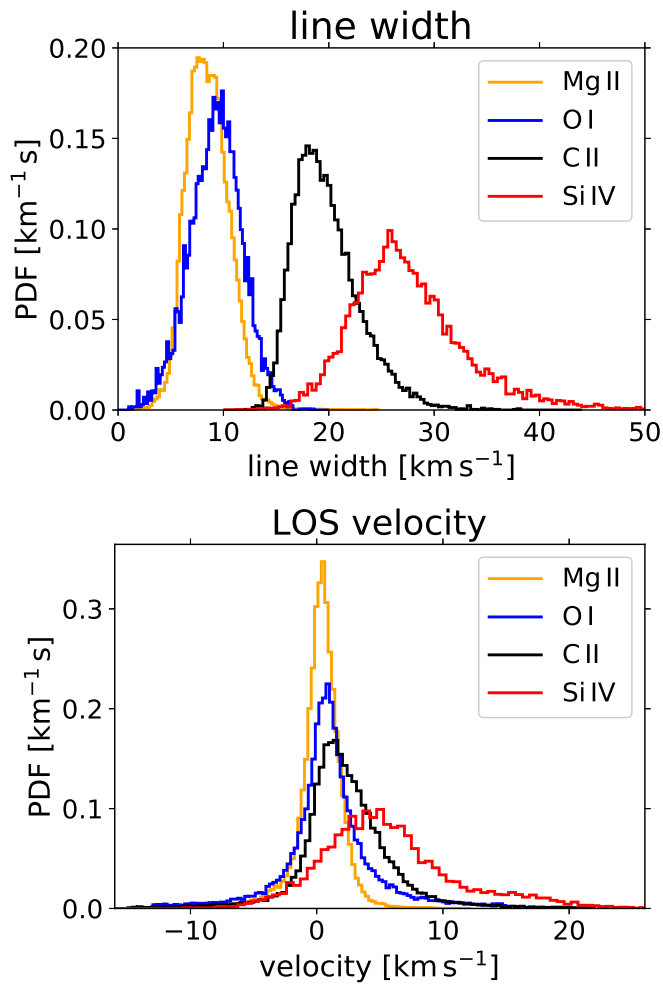}}
\caption{Histograms of the line widths (top) and velocity (bottom) in the LB sample. 
The line width is equal to $2\,\sigma$ where $\sigma$ is the Gaussian width. 
For a description of the velocity calibration, see Sect.\,$2.2$. Positive velocity corresponds to downflow.}
\label{fig:hist_lw}
\end{figure}
\subsection{Far ultraviolet spectrum}\label{sect:fuv_spectrum}
The far-Ultraviolet (FUV) spectrograph often records four wavelength bands: the \ion{C}{ii} lines at $133.6$\,nm;   
the \ion{O}{i} and \ion{C}{i} lines at $135.6$\,nm;  the \ion{Si}{iv} line at $139.3$\,nm; and  
the \ion{O}{iv} line at $140.1$\,nm and the \ion{Si}{iv} line at $140.3$\,nm. 
The chromospheric \ion{O}{i} line at $135.56$\,nm  forms at a temperature range of $7-16\times\,10^3$\,K,
comparable to the formation temperature of the \ion{Mg}{ii}\,k line.
The \ion{O}{i}\,$135.56$\,nm and the \ion{C}{i}\,$135.584$\,nm lines are fitted together. 
The \ion{O}{i} line is used to construct a velocity modulation curve 
to correct for the orbital motion of the satellite in the FUV channel (the procedure is identical to the NUV case). 
The \ion{C}{ii} line pair, the \ion{Si}{iv} lines, and the \ion{O}{iv} lines form in the TR. 
{ Although the \ion{C}{ii} ionization fraction peaks at $2.5\times\,10^4$\,K \citep{rathore_carlsson_2015}, 
the \ion{Si}{iv}\,$140.277$\,nm line forms at $\log\rm{T[K]}\approx\,4.8$ \citep{peter_2001}
and the \ion{O}{iv} lines form at $\log\rm{T[K]}\approx\,5.4$ \citep[][ their Table~V]{arnaud_rothenflug_1985}. 
 Table~\ref{tab:lines} lists the formation temperature, rest wavelength, and excitation potential of selected spectral
lines in the IRIS spectra.  
While the \ion{C}{ii} line pair are often optically thick,
the \ion{Si}{iv} and \ion{O}{iv} lines only show the double reversal pattern in flaring regions, for example. 
From the two \ion{Si}{iv} lines, we only discuss the \ion{Si}{iv}\,$140.277$\,nm line
in this paper as the \ion{Si}{iv}\,$139.376$\,nm line is blended with a \ion{Ni}{ii} line and is not always recorded. 
A spectral synthesis of these lines \citep[using CHIANTI,][]{landi_etal_2013_chianti} shows 
that the \ion{C}{ii}, the \ion{Si}{iv}, and the \ion{O}{iv} lines form at a limited temperature range. 
The spectral dispersion in the FUV channel was either $1.296$ or $2.592$\,pm 
corresponding to a velocity dispersion of $2.8$ or $5.6$ km\,s$^{-1}$ per pixel, respectively.

\paragraph{Data analysis:}
We developed a new data reduction package for the IRIS spectra that analyzes
the major spectral lines \citep{mosic_2017}. 
The program\footnote{It is also available at {http://github.com/reza35/mosic}} 
was tested on a variety of datasets and is available in the SolarSoft \citep{solarsoft_1998}. 
For the NUV data, it performs a spectral analysis and a Gaussian fit to the core of
the \ion{Mg}{ii}\,h and k lines and a few nearby lines. 
For the FUV spectral channels, a multi-line Gaussian fit is performed. Details of
the fitting procedure is explained in the following. 
In the fit procedure the observational errors,
which are a joint action of the photon and thermal noise, are taken into account.
We used the hairlines to align the NUV and FUV data (Fig.~\ref{fig:mask}).

\subsection{Near-ultraviolet data}
We define the standard parameters of double-reversal profiles of the \ion{Mg}{ii}\,h and k lines in a similar way to 
the \ion{Ca}{ii}\,H and K lines \citep{cram_dame_83, lites_etal_93, reza_etal_3, beck_etal_08}. 
The core of the \ion{Mg}{ii}\,h and k lines forms higher than the 
\ion{Ca}{ii}\,H and K lines, due to a larger opacity \citep{leenaarts_etal_2013a}. 
While the $\rm{k}_1$ minima form at about the temperature minimum,
the $\rm{k}_2$ and $\rm{k}_3$ form above $1$\,Mm \citep{leenaarts_etal_2013b}. 
At first we fit a single-Gaussian function to the Mg\,{\sc ii}\,h and k lines 
between k$_{\mathrm{1}}$ (h$_{\mathrm{1}}$) minima after removing the wing slope. 
After that, we proceed with a double- and a triple-Gaussian fit. 
A guided multi-Gaussian fitting method was previously presented by several authors \citep{peter_2000, tian_etal_2011}.
In this approach, the result of each step is used to construct an initial guess for
the next step that incorporates a more complex model. 
We use the MPFIT package for the Gaussian fits  discussed in this paper \citep{mpfit} .

\begin{figure*}
\resizebox{\hsize}{!}{ \includegraphics*{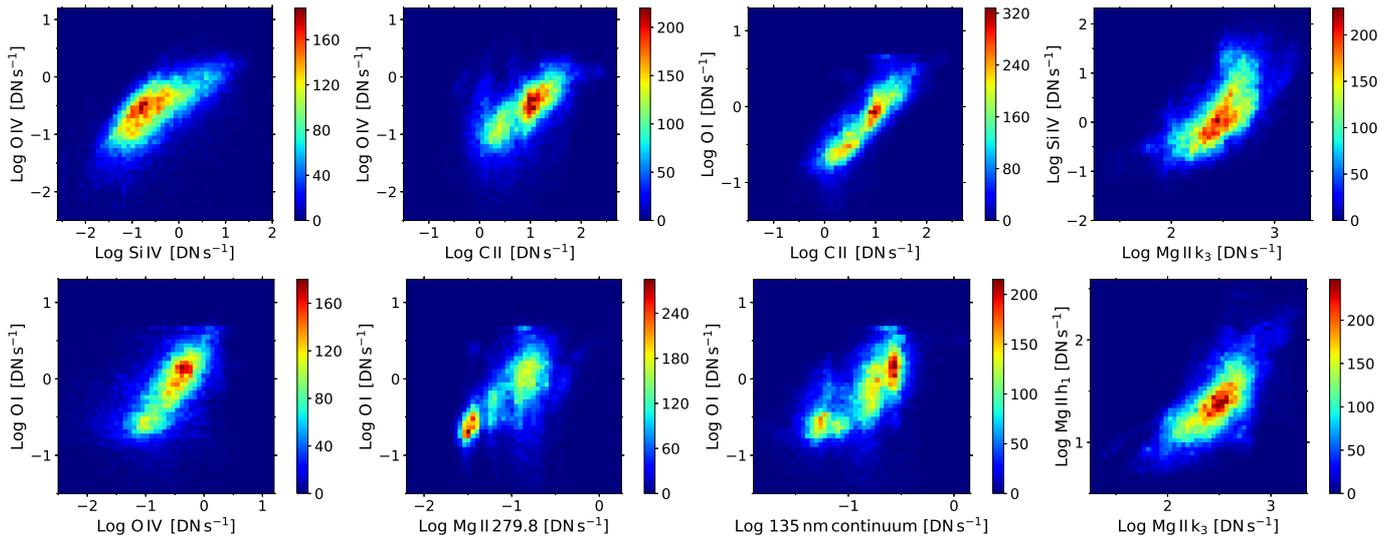}}
\caption{Two-dimensional density plots of the intensity parameters in the FUV and NUV spectra. 
  Color bars denote the density of points. The h$_1$ intensity and the
  line-core intensity of the subordinate \ion{Mg}{ii}\,$297.80$\,nm line are the result of  a profile analysis
  while the rest are retrieved from  the Gaussian fits. A positive correlation is seen in all panels.}
\label{fig:int_scat}
\end{figure*}
\begin{figure*}
\resizebox{\hsize}{!}{ \includegraphics*{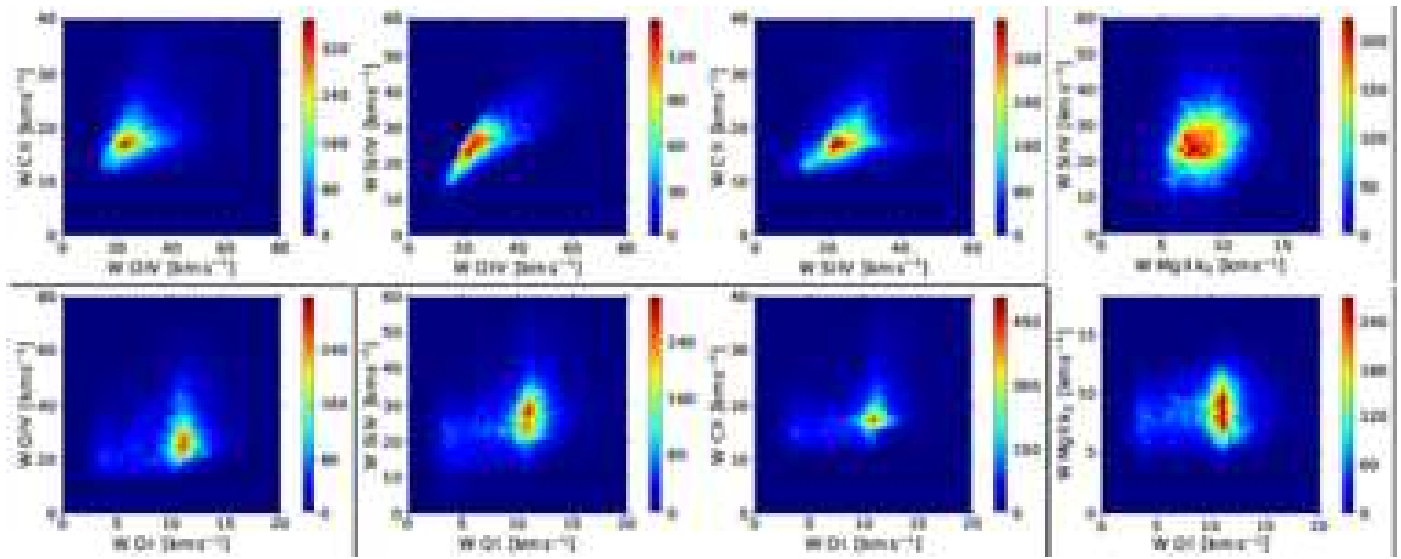}}
  \caption{Two-dimensional density plots of the line widths ($W\,=2\,\sigma$) in different lines.
    There is a correlation between the width of the FUV lines (\ion{C}{ii}, \ion{Si}{IV}, \ion{O}{iv}), 
    while there is no significant correlation between the chromospheric and the FUV lines.}
\label{fig:ww}
\end{figure*}
\subsection{The FUV data}\label{sec:si}
We used a multi-line multi-Gaussian fitting approach to simultaneously fit all spectral lines 
between $139$ and $141$\,nm listed in Table\,\ref{tab:lines}.   
The fitting procedure is similar to the \ion{Mg}{ii} case where we gradually
increased the number of the free parameters to have a better fit 
and evaluate the results using the reduced chi-square~\citep{bevington_robinson_1992}. 
The same technique was used to fit the \ion{O}{i}\,$135.56$\,nm and \ion{C}{ii}\,$133.45$\,nm lines.
The \ion{O}{i}\,$135.56$\,nm line forms at a height of about $1.0-1.5$\,Mm and is always optically thin \citep{lin_carlsson_2015}.
The \ion{C}{i}\,$135.58$\,nm line in the red wing of the  \ion{O}{i} line is
usually weaker, but can be stronger in flares \citep{cheng_etal_1980}. 
For the \ion{Si}{iv}\,$140.3$\,nm line, we also performed a double-Gaussian fit. 
The two Gaussians do not have similar properties: one has a larger amplitude, a narrower line, 
and a smaller shift (core component), while the other usually has a smaller amplitude, a broader line, 
and a larger shift (tail component). The core and tail components retrieved during this process are
discussed in Sect.~\ref{sec:intensity}.

\begin{table}
\centering  
\caption{ Atomic parameters of selected spectral lines. $\lambda$ is the observed wavelength in air. 
$\chi$ stands for the excitation potential of the lower level. 
The formation temperatures are from \citet{sutherland_dopita_1993}. 
The main spectral lines discussed in the paper are marked with an asterisk.}
\label{tab:lines}
\begin{tabular}{l l l l l} \hline\rule[-1.5mm]{-1mm}{5mm}
Line           &  $\lambda$\,[nm] & $\chi$\,[eV] &  $\log$\,T\,[K] & reference\\\hline\rule[-1.5mm]{-1mm}{5mm}
C\,{\sc ii}{$\ast$}  &  $133.4532$     &  $1.264$   &  $4.20-4.65$  &  1, 2\\
Ni\,{\sc ii}   &  $133.5201$     &  $0.187$   &  $4.00-4.25$  &  3, 4\\
C\,{\sc ii}    &  $133.5708$     &  $1.264$   &  $4.20-4.65$  &  1, 2\\
O\,{\sc i}$\ast$  &  $135.5598$  &  $0.000$   &  $4.00-4.15$  &  1, 5\\
C\,{\sc i}        &  $135.5844$  &  $1.264$   &  $4.00-4.10$  &  1, 2\\\hline\rule[-1.5mm]{-1mm}{5mm}
Si\,{\sc iv}   &  $139.376$      &  $0.000$   &  $4.05-4.25$  &  3, 6\\
Ni\,{\sc ii}   &  $139.902$      &  $0.187$   &  $4.00-4.25$  &  3, 4\\
O\,{\sc iv}    &  $139.978$      &  $0.000$   &  $4.05-4.25$  &  3, 6\\
Fe\,{\sc ii}   &  $139.9957$     &  $2.778$   &  $4.05-4.25$  &  1, 7\\
O\,{\sc iv}$\ast$    &  $140.115$      &  $0.048$   &  $5.15-5.30$  &  2, 5\\
S\,{\sc i}     &  $140.151$      &  $0.000$   &  $< \,4.00$   &  1, 8\\
Si\,{\sc iv}$\ast$   &  $140.277$      &  $0.000$   &  $4.75-4.85$  &  1, 9\\
Fe\,{\sc ii}   &  $140.5602$     &  $0.232$   &  $4.05-4.25$  &  1, 7\\
S\,{\sc iv}    &  $140.481$      &  $0.000$   &  $4.95-5.05$  &  1, 9\\
O\,{\sc iv}    &  $140.481$      &  $0.048$   &  $5.15-5.30$  &  3, 6\\
S\,{\sc iv}    &  $140.608$      &  $0.117$   &  $4.95-5.05$  &  1, 9\\
O\,{\sc iv}    &  $140.738$      &  $0.048$   &  $5.15-5.30$  &  1, 5\\\hline\rule[-1.5mm]{-1mm}{5mm}
Mg\,{\sc ii}\,k$\ast$ & $279.553$      &  $0.000$   &  $4.0-4.20$   &  1, 10\\
Mg\,{\sc ii}   &  $279.800$      &  $4.433$   &  $4.0-4.20$   &  1, 10\\
Ni\,{\sc i}    &  $279.865$      &  $0.110$   &  $< \,4.00$   &  1, 11\\ 
Fe\,{\sc i}    &  $283.2435$     &  $0.958$   &  $< \,4.00$   &  1, 7\\\hline
\end{tabular}
\tablebib{
  1-~\citet{nist}; 2-~\citet{moore_1970c}; 3-~\citet{vald}; 4-~\citet{K03}, 5-~\citet{moore_1993}; 6-~\citet{K11};
  7-~\citet{nave_etal94};
  8-~\citet{kaufman_1982}; 9-~\citet{toresson_1960}, 10-~\citet{risberg_1955}; 11-~\citet{mn_i_1975}}
\end{table}

The UV continuum at $135$\,nm forms beyond the temperature minimum \citep{judge_2006}. 
To estimate the UV continuum,
we computed the median of a subset of continuum windows excluding the $1$\,\% outliers, 
which was then used as an initial guess in the Gaussian fits. This 
resulted in a fairly clean continuum map but sometimes there are a few warm or hot pixels. 
There is an offset between the upper and lower halves of the chip.
This offset is due to simultaneous readout of two cameras and is present in some datasets. 
In addition in some datasets, the dark subtraction in the calibration procedure creates a negative UV continuum in the
level\,$2$ data. We correct the negative continuum values in the maps 
by lifting the lower limit of the continuum distribution to have a positive minimum. 
A new IRIS technical document will address these issues.

Unlike photospheric spectral lines which have a well-known convective blueshift pattern and have an
average velocity in quiet Sun areas close to zero \citep{doerr_phd},
the average quiet Sun TR line profiles show significant Doppler shifts \citep{peter_judge_1999}. 
For the \ion{Si}{iv} line, we compared the line positions with the mean position of the \ion{S}{i}\,$140.151$\,nm line
(assuming this line to be at rest),
while for the \ion{C}{ii} line the nearby \ion{Ni}{ii}\,$133.520$\,nm was used as a reference,
again assuming that the mean profile of this line has zero Doppler shift.
The accuracy of these assumptions is about $1-2$\,km\,s$^{-1}$.

\begin{figure}
\resizebox{\hsize}{!}{\includegraphics*{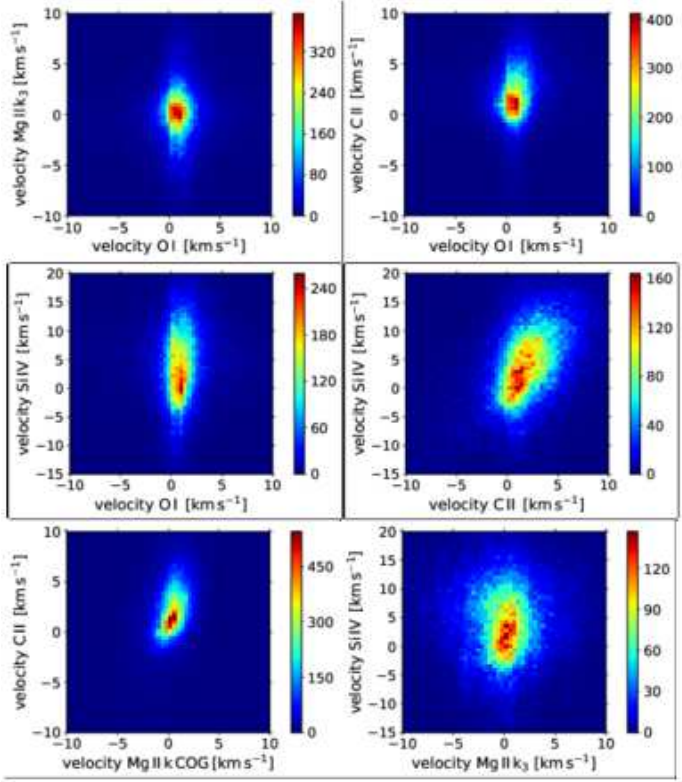}}
  \caption{Same as Fig.~\ref{fig:int_scat} but for the Doppler shifts.}
\label{fig:velo}
\end{figure}
\begin{figure}
\resizebox{\hsize}{!}{\includegraphics*{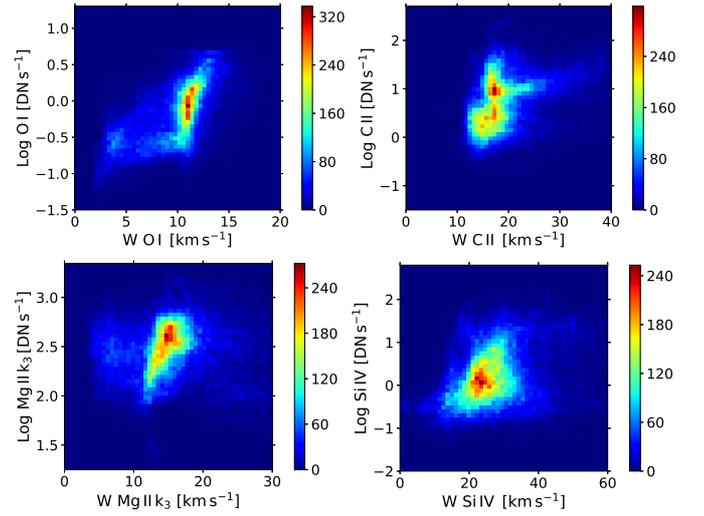}}
\caption{Two-dimensional density plots of the logarithm of the line intensity versus the line width.}
\label{fig:intensity_width}
\end{figure}
\section{Results}\label{sec:result}
The analysis of $60$ LBs resulted in a sample of several NUV and FUV line parameters.
The LB sample consists of $42,436$ entries for each parameter. 
In the following, we mainly discuss the peak intensity, the line width, and the velocity of the \ion{Mg}{ii}\,k,
\ion{O}{i}, \ion{C}{ii}, and \ion{Si}{iv} lines. 
Light bridges are clearly seen in the core of the \ion{Mg}{ii}\,k and \ion{Si}{iv} lines. 
While the IRIS scanning slit clearly leaves the pattern of umbral flashes in the maps of the magnesium line core, 
the \ion{Mg}{ii}\,k$_{\rm{3}}$ intensities do not show umbral flashes in the LBs.

Figure~\ref{fig:hist_int} shows distributions of the continuum and line intensities in the LB sample.
It is seen that LBs have highest intensity in the \ion{Mg}{ii}\,k line, an order of
magnitude higher than the corresponding $280$\,nm continuum. 
The median LB intensity in the $280$\,nm continuum and 
the \ion{Mg}{ii}\,k line (k$_{\rm{3}}$) are $23.1$ and $306.4$\,DN\,s$^{-1}$, respectively. 
Excluding $5$\% outliers in the histograms, a probability interval covering $90$\% of all entries
for the $280$\,nm and the \ion{Mg}{ii}\,k line corresponds to $[6.4,\,72.6]$ and $[97.0,\,752.5]$\,DN\,s$^{-1}$, respectively. 
The \ion{O}{i}\,$135.560$\,nm line has a median intensity of $0.7$\,DN\,s$^{-1}$
with a $90$\,\% probability interval of $[0.2,\,3.9]$\,DN\,s$^{-1}$, which  
is comparable to the \ion{O}{iv}\,$140.115$\,nm forbidden line with an average amplitude of
$0.2$\,DN\,s$^{-1}$ and a $90$\,\% probability interval of $[0.0,\,1.2]$\,DN\,s$^{-1}$. 
The ratio of the line-core intensity of the \ion{C}{i}\,$135.58$ and \ion{O}{i}\,$135.56$\,nm lines is distinctly
different in the LB sample than in the quiet Sun sample. Excluding all pixels that do not have a clear signal in both lines, 
the median ratio in the LB sample is about $0.5$, while it is $0.3$ in the quiet Sun sample.  
The intensities of the \ion{C}{ii} and \ion{Si}{iv} lines are $5.8$ and $1.3$\,DN\,s$^{-1}$, respectively,
about an order of magnitude higher than the \ion{O}{i} and the \ion{O}{iv} lines.
The $90$\,\% probability interval of the \ion{C}{ii} line corresponds to $[0.8,\,42.1]$\,DN\,s$^{-1}$,
of the same order of magnitude as the \ion{Si}{iv} range ($[0.2,\,23.0]$\,DN\,s$^{-1}$).
For a double-Gaussian fit to the \ion{Si}{iv} line, the core and tail components have
a median intensity of $1.3$ and $0.2$\,DN\,s$^{-1}$, respectively. 
The UV continuum can be measured in the $133$, $135$, or $140$\,nm wavelength range, and the resulting maps are fairly comparable. 
The UV continuum discussed below is retrieved from the $135$\,nm wavelength range.
It has the smallest intensity consistent with the fact that all the studied FUV lines are in emission:
the UV continuum has a median intensity of $0.2$\,DN\,s$^{-1}$ with a $90$\,\% probability interval of $[0.0,\,0.3]$\,DN\,s$^{-1}$.

\begin{figure*}
\resizebox{\hsize}{!}{\includegraphics*{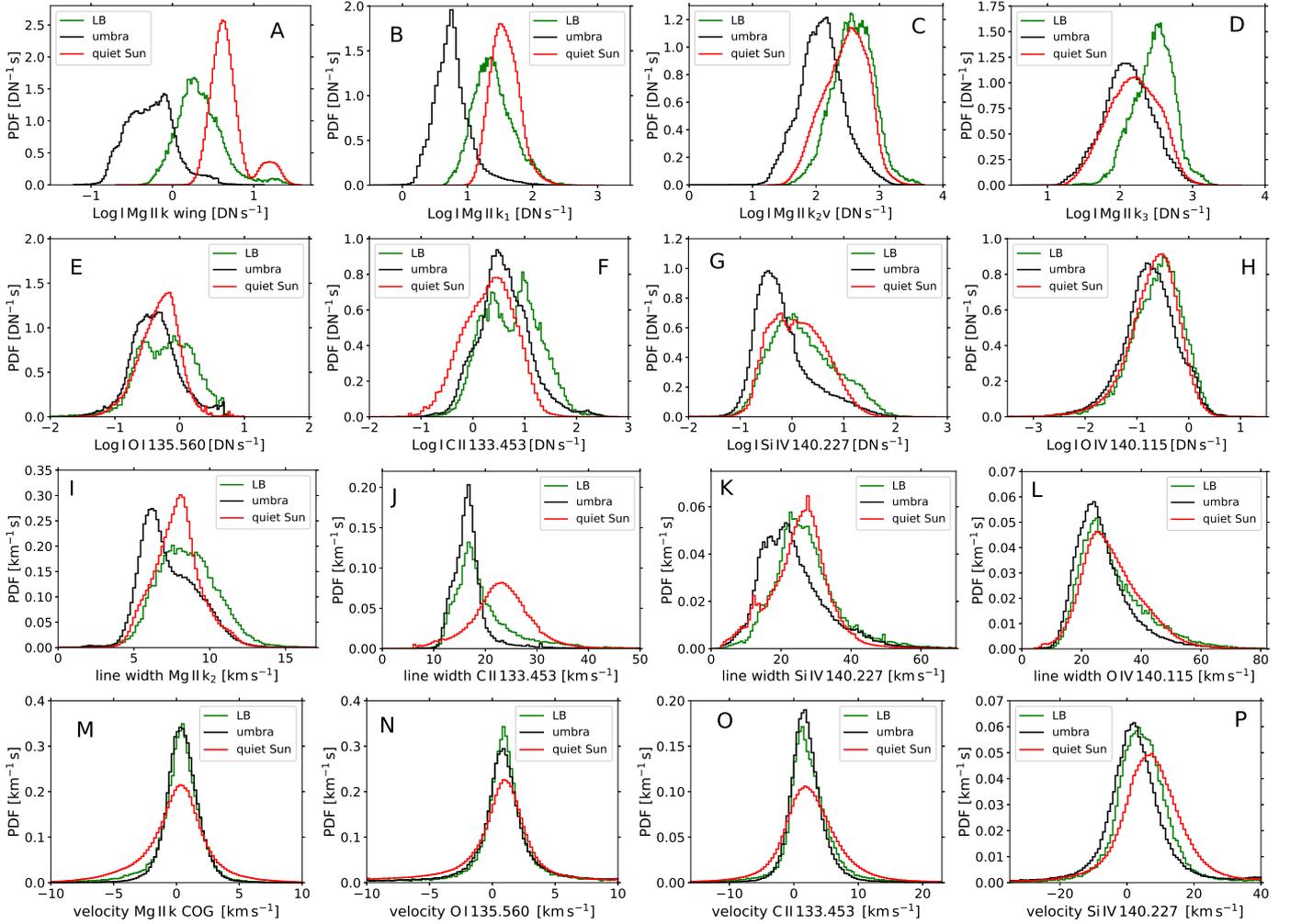}}
  \caption{Distribution of the intensity, line width, and velocity in the LB sample (green) compared to the umbra (black) and
    quiet Sun (red) samples. %
    The quiet Sun area  surrounding sunspots includes some plage regions seen as a secondary population in the photosphere (panel \emph{A}).
    Histograms of the intensity parameters are shown in panels \emph{A-H}, while panels \emph{I-L} and \emph{M-P} show histograms
    of the line widths and  velocities, respectively. 
    The line width of \ion{Mg}{ii}\,k$_{\rm{2}}$ shown on panel (\emph{I}) marks the average width of the two Gaussians
    used to fit the line. All intensity parameters (panels \emph{A-H}) were normalized to the corresponding integration times.
    Positive velocity denotes redshift.}
\label{fig:hist}
\end{figure*}
The top panel of Fig.~\ref{fig:hist_lw} shows the line width distribution in the LB sample. 
The line width is measured at $1/e$ of the peak intensity  and is equal to $2\,\sigma$, where $\sigma$ is the Gaussian width. 
The line width of the \ion{O}{i} and \ion{Mg}{ii}\,k lines (mean of the k$_{\rm{2}}$ emission widths)
are comparable ($8.4$ and $10.6$\,km\,s$^{-1}$), as was 
expected since these lines mainly form in a similar temperature range.
The \ion{O}{i} line width of $90$\,\% of all entries in the LB sample falls between $[4.0,\,14.5]$\,km\,s$^{-1}$,
which spans a broader range than the \ion{Mg}{ii}\,k$_{\rm{2}}$ range ($[5.6,\,11.8]$\,km\,s$^{-1}$). 
The \ion{C}{ii} line has a median width of $18.0$\,km\,s$^{-1}$ and the $90$\,\% probability interval was 
$[13.0,\,32.2]$\,km\,s$^{-1}$. 
The single-Gaussian fit to the \ion{Si}{iv} line has a median width of $22.0$\,km\,s$^{-1}$ and
the $90$\,\% probability interval covers a  broader range than the \ion{C}{ii} line ($[13.4,\,44.9]$\,km\,s$^{-1}$).
Applying a double-Gaussian fit, the core and tail
components have a median width of $20.0\,\,$and $29.8$\,km\,s$^{-1}$  and the corresponding $90$\,\%  probability intervals are
$[12.9,\,40.1]$ and $[12.8,\,85.8]$\,km\,s$^{-1}$, respectively. 
A single-Gaussian fit to the \ion{O}{iv}\,$140.115$\,nm line (Table~\ref{tab:lines}) yields  
a median width of  $25.3$\,km\,s$^{-1}$ and the $90$\,\% probability interval corresponds to $[15.8,\,45.1]$\,km\,s$^{-1}$. 
The extreme largest line widths resemble flaring regions. 
The FUV observed lines are significantly wider than the sound speed at the corresponding 
formation temperature \citep[$4.3$\,km\,s$^{-1}$ for the \ion{C}{ii} and \ion{Si}{iv} lines,][]{peter_2001}. 
Distribution of the velocity in the LB sample is shown in the bottom panel of Fig.~\ref{fig:hist_lw}. 
Light bridges are redshifted, almost in all spectral lines. The blueshifts in the \ion{O}{i} and \ion{Mg}{ii} lines
are comparable to the redshifts. The median velocity in these two lines are $1.3$ and $0.4$\,km\,s$^{-1}$, and the 
$90$\,\% probability intervals correspond to $[-4.1,\,5.7]$ and $[-3.1,\,2.7]$\,km\,s$^{-1}$, respectively. 
The \ion{C}{ii} line is systematically redshifted ($2.0$\,km\,s$^{-1}$) with a $90$\,\% probability interval of $[-3.2,\,8.7]$\,km\,s$^{-1}$.
The \ion{Si}{iv} line has the largest redshift and $90$\,\%  probability interval ($4.3$,\,$[-9.1,\,22.9]$\,km\,s$^{-1}$).

Figure~\ref{fig:int_scat} shows two-dimensional density plots of the line intensities in the NUV and FUV spectra. 
From the $280$\,nm photospheric continuum to the \ion{O}{iv} line core, 
there is a significant spatial correlation between all pairs of the line and continuum intensities.  
Noteworthy is the fact that even the UV continuum and the subordinate \ion{Mg}{ii}\,$279.80$\,nm line-core intensity which have 
weak signals clearly show the correlation to the other lines.
In most panels, the parameters on either axis were retrieved from the analysis of a different spectral band.

Density plots of the line width (at $1/e$ of the peak amplitude) for all lines in our study are shown in Fig.~\ref{fig:ww}. 
The width of the \ion{C}{ii} line has a significant correlation only with the \ion{Si}{iv} and \ion{O}{iv} lines.
The width of the k$_{\rm{2}}$ peaks corresponds to the average width of the two Gaussians used to fit the line and 
does not represent the distance of the two emission peaks. 
The \ion{Mg}{ii}\,k$_{\rm{2}}$ width does not scale with the width of the FUV lines.
The same is true for the width of the \ion{O}{i} line (Fig.~\ref{fig:ww}, bottom panels):
the width of these two lines changes in a restricted range. 
The tail toward the very small line width in the \ion{O}{i} panels is perhaps due to noise. 
The density plot of the \ion{Mg}{ii}\,k line versus the \ion{O}{iv} line (not shown here) is very
similar to that for the \ion{Si}{iv} line (top right panel).
In summary, we found evidence of a correlation between the width of the TR lines while 
there is no significant correlation between the width of the TR and chromospheric lines.

Density plots of the Doppler shifts are shown in Fig.~\ref{fig:velo}. 
All lines are dominated by redshifts, which is also clear in the velocity
histograms (Fig.~\ref{fig:hist_lw}, bottom panel).
The velocity of the \ion{Si}{iv} and \ion{C}{ii} lines span a much broader range than the chromospheric lines.
There is a significant correlation between 
the velocity in the \ion{Mg}{ii}\,k center of gravity and the \ion{C}{ii} lines as well as between the \ion{C}{ii} and \ion{Si}{iv} lines.
The relation between the magnesium k$_{\rm{3}}$ and the silicon line-core velocities shows some trend with a large scatter.
The velocity of the \ion{O}{i} line does not clearly show a trend with other lines. 
There is also a positive correlation between the intensity and the line width in all the lines (Fig.~\ref{fig:intensity_width}). 
Although the line intensities span several orders of magnitude, the line widths changes within a factor of two. 
The scatter increases toward the hotter lines.

A comparison of the histograms of the LB sample with the umbra and quiet Sun samples is displayed in Fig.~\ref{fig:hist}.
The wing intensity of the \ion{Mg}{ii}\,k line in quiet Sun is brighter than the LBs and umbra (panels \emph{A}). 
The same is true for the continuum intensity at the $280$\,nm (not shown here). 
As seen in panels \emph{B-D}, LBs are one order of magnitude brighter than umbra in the \ion{Mg}{ii}\,k line. 
In panels \emph{D-G}, the LBs are brighter than the quiet Sun in the \ion{Mg}{ii}\,k, \ion{O}{i}, \ion{C}{ii}, and \ion{Si}{iv} lines. 
The intensity enhancement compared to umbra in the TR lines is smaller than the \ion{Mg}{ii}\,k 
line (panels \emph{E-H}). A comparison of the line width is shown in panels \emph{I-L}.
Light bridge profiles are either significantly wider than umbra 
(\ion{Mg}{ii}\,k and \ion{C}{ii} lines, panels \emph{I-J}) 
or have an extended tail toward very broad lines (panels \emph{K-L}).} 
Although LBs are dominated by redshifts, the velocity distribution are blueshifted in 
the LBs and umbra samples compared to the quiet Sun sample (panels \emph{O} and \emph{P}). 
The chromospheric velocities of our LBs shown in panels \emph{M} and \emph{N} are close to umbra.

\begin{figure}
\resizebox{\hsize}{!}{\includegraphics*{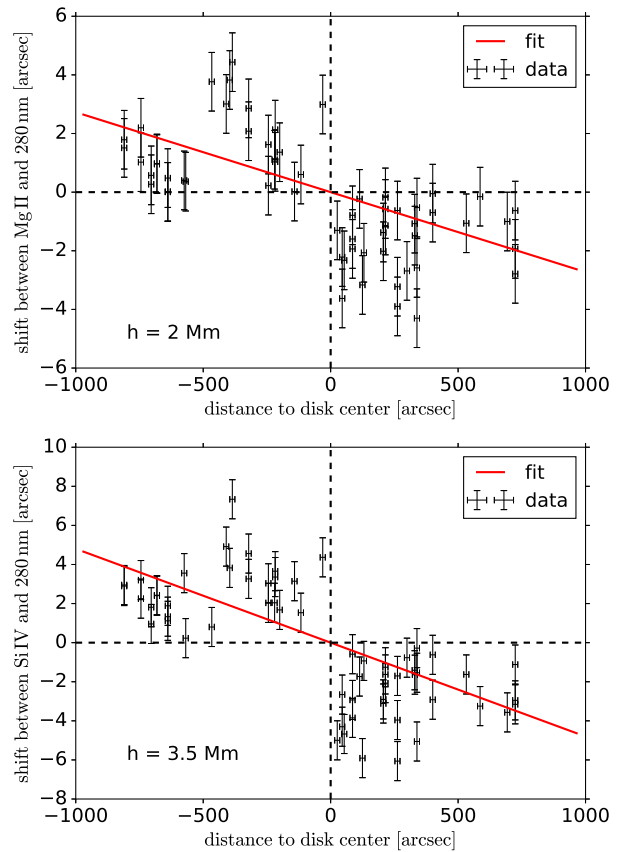}}
\caption{Scatter plot of the displacement between center of mass of the LBs in the $280$\,nm continuum
  channel and the \ion{Mg}{ii} and \ion{Si}{iv} lines. The $x$-axis represents the radial distance
  to  the disk center in the heliographic coordinate. 
  The red line shows a linear fit passing through the origin. The estimated altitude of each line
as the displacement value at  one solar radius is shown in each panel.}
\label{fig:shift}
\end{figure}

Light bridges are systematically displaced in different spectral lines
such that high forming lines are closer to the solar limb. To estimate this projection effect, we used the $280$\,nm
continuum, the \ion{Mg}{ii}\,k, and the \ion{Si}{iv} masks to find the center of mass
of the LBs in different channels. The displacement along the radial direction 
was then compared to the distance to the disk center (Fig.~\ref{fig:shift}). 
The displacement and the $x$ coordinate have opposite signs so a LB is closer
to the disk center in the photosphere than in the TR. 
The amount of the displacement at one solar radius
in the \ion{Mg}{ii}\,k\,\,/\,\,\ion{Si}{iv} maps is about $2.0\,/\,3.5$\,Mm which corresponds to the 
formation height of the chromosphere\,\,/\,\,TR.
The deviations compared to the linear fit can be due to a mismatch of the
exact shape of the LBs, its extension, or a real height difference between different LBs in the sample.\vspace{-0.5cm}

\section{Discussion}\label{sec:discussion}
\subsection{Vertical extent}
{We measured a systematic displacement between the center of mass position of the LBs 
in the TR and chromosphere compared to the photosphere. 
The LBs in the photospheric continuum are closer
to the disk center than in the \ion{Si}{iv} and \ion{Mg}{ii} maps 
which hints at the three-dimensional structure of LBs in the solar atmosphere. 
The scatter in Fig.~\ref{fig:shift} close to the disk center is significant as 
the absolute amount of the shift is smaller close to the disk center than close to the limb.
Therefore the measured shifts close to the disk center are more prone to the
noise (the shifts are partially affected by the mismatch of the exact shape of the LBs in different layers). 
This displacement is partially responsible for the scatter in the density plots (Fig.~\ref{fig:int_scat}): 
there is a larger scatter between two parameters that form at very different
temperatures than those that form at a comparable temperature. 
This systematic displacement also indicates that LBs to some extent are
coherent structures in an extended height range. 
The fact that each IRIS scan was recorded over approximately one hour does not necessarily imply that 
the measured emission enhancements associated with LBs should have a stationary nature. 
A quasi-stationary bright LB can be sustained through the transient events provided
that the brightenings happen with a high enough frequency as
suggested by \citet{toriumi_etal_2015}, which is consistent with observations of
LBs with IRIS filtergrams reported by \citet{bharti_2015}
and \citet{yang_etal_2015}. These authors found frequent small-scale jets associated with LBs.
Although there is no temporal evolution in the LB sample, our results remain qualitatively 
consistent with an intermittent supply of the energy and momentum to LBs.

The presence of a systematic LB signal in the chromosphere and TR is not consistent with a cups-shaped 
magnetic configuration \citep{leka1997, jurcak_etal_2006,lagg_etal_2014, felipe_etal_2016}.
According to this model, the field lines associated with LBs merge in the upper photosphere. 
Therefore, in the chromosphere and TR, the LB field lines are not distinguishable from the umbra.
This is inconsistent with a systematic observation of LBs in the upper atmosphere. 


\subsection{Intensity}\label{sec:intensity}
Light bridges show an excess intensity that can be more than one order of magnitude higher than the 
nearby umbra. The intensity of the UV continuum (at $135$\,nm) is about $0.2$\,\,DN\,s$^{-1}$ and is significantly more
prone to noise than line-core intensities. While not true in all cases, there are many LBs in which the 
UV continuum maps show  a pattern of the photospheric LB. 
The UV continuum at $135$\,mm mimics the local heatings at $\succsim\!0.5$Mm. 
We also note that the subordinate \ion{Mg}{ii}\,$279.80$\,nm line 
in the red wing of the  \ion{Mg}{ii}\,k line shows {\em emission} in several  LBs, distinct from the observed 
absorption line in the photosphere and plage \citep{carlsson_leenaarts_depon_2015}.  
These findings provide evidence for an energy deposit in the low chromosphere.

The blue wing of the \ion{Mg}{ii}\,k line (Fig.~\ref{fig:hist}, panel \emph{A})
has a similar intensity distribution to that of the $280$\,nm continuum (Fig.~\ref{fig:hist_int}).
The magnesium k$_{\rm{2}}$ and k$_{\rm{3}}$ bands have an intensity that is higher 
than the co-spatial and co-temporal continuum by more than one
order of magnitude (Fig.~\ref{fig:hist}, panels \emph{C-D}), which 
indicates that the radiative processes cannot provide enough energy
and there must be some sort of chromospheric heating that sustains the \ion{Mg}{ii}\,k line-core emission. 
The asymmetry between the violet and red emission peaks in the 
LBs is comparable to quiet Sun (k$_{\rm{2v}}$ is slightly stronger than k$_{\rm{2r}}$). 
The emission strength, the ratio between the k$_{\rm{2v}}$ and k$_{\rm{3}}$, is between umbra and quiet Sun.  
Unlike nearby umbral profiles which have a single lobe, LB profiles show the
typical double-reversal pattern, while sometimes the two lobes merge and form a very asymmetric line profile. 
The \ion{Mg}{ii}\,k line profiles are distinctly different from that in an umbra or a plage:
the lines are broader than umbra, but not as broad as plage (particularly not as broad in the k$_{\rm{1}}$ level).
The line width remains comparable to quiet Sun, although it is usually smaller. 
This has two implications: First, the LB temperature stratification in a one-dimensional model atmosphere
should significantly deviate from an umbral (or a plage) atmospheres. 
Second, the umbral scattered light does not play an important role in LB profiles as 
there is a clear distinction between the LB and umbra profiles (umbra profiles are narrower, 
have a single emission peak, and have an intensity up to three orders of magnitude smaller than LB profiles). 
The strong correlation between the intensity in different continuum and lines is
comparable to quiet Sun areas \citep{doschek_etal_2004}. Our results remain consistent
with \citet{chae_etal_1998b}, who suggested that the lower to mid TR comprises coherent structures.

As discussed by \cite{peter_2000, peter_2010}, many spectral lines in the TR including the \ion{Si}{iv}\,$140.3$\,nm line
show a tail component in the network regions while a single-Gaussian fit is enough to fit the observed profile in inter-network. 
The double-Gaussian fit of our LBs (Sect.~\ref{sec:si}) resulted in an average tail component 
that is about $14\%$ of the total intensity (tail plus core). 
\citet{peter_2001} studied the core and tail components in quiet Sun network and
reported a similar fraction ($14.8\%$). 
The average line width measured in our LBs are, however, significantly smaller than that reported by \citet{peter_2001}
for networks ($22/20/30$ vs. $28/25/44$ km\,s$^{-1}$ for the single-Gaussian, and the core and tail components of the double Gaussian). 
The higher spatial resolution of IRIS data compared to SUMER implies that the
difference can be partially due to the instrumental effects 
(a line profile emerging from a larger resolution element has a larger width due to 
the velocity dispersion compared to that originating from a smaller area). 
Further studies of networks using IRIS data facilitates a more accurate comparison.

\cite{yang_etal_2015} found LB traces in the IRIS $133$\,nm slitjaw images co-temporal and co-spatial with 
high-resolution \ion{TiO}{} filtergrams from ground-based observations. 
IRIS filtergrams at $133$ and $140$\,nm have a mixed contribution from the core and continuum of the FUV lines.
Unlike photospheric and chromospheric filtergrams in the visible/near-infrared in which 
spectral line-core intensities are within a factor $2-10$ of the nearby continuum, a line-core intensity 
in the UV lines can be a factor of $10^4$ higher than the nearby continuum. 
Therefore, even if the continuum wavelength range 
covers a band that is hundred times larger than the width of the strong FUV line in the filter passband
\citep[$0.4$\,nm for the \ion{Mg}{ii} filter and $4$\,nm for the $133/140$\,nm filters,][]{iris_2013}, the  
contribution of the line to the integrated filtergram signal can exceed that from the UV continuum by a large factor.  
As a result, in active regions, the line mostly outweights the continuum signal in the IRIS filtergrams, while in quiet Sun 
the filtergrams represent the UV continuum. In contrast, the line intensities from IRIS spectra  presented here 
have no contribution from the line wing or nearby continuum.
Similarly, the UV continuum in our data shows a pure continuum signal that is not contaminated through the wings of spectral lines.
This distinction is absent in the IRIS filtergrams of LBs presented by \citet{yang_etal_2015} and \citet{bharti_2015}.

\subsection{Line width}
The line widths in the chromosphere and TR are primarily nonthermal as the
turbulence velocity increases toward the upper umbra atmosphere \citep{kneer_mattig_1978, lites_skumanich_1982};  
they are also nonthermal in a quiet Sun atmosphere \citep{val81}.
As seen in Fig.~\ref{fig:hist_lw}, the line width increases with the formation temperature, which 
is in accordance with the general understanding that the maximum nonthermal broadening happens
at about $3\times10^5$K in quiet Sun, beyond the temperature range discussed here \citep{dere_mason_1993, peter_2001}. 
The increased broadening of the chromospheric and TR lines in the LBs compared to umbra 
provides evidence for an energy deposit in the atmosphere. 
Among different line broadening mechanisms, we note a few of them relevant to the discussed spectral lines.
A line can be broadened by the turbulence in the atmosphere \citep{afy}, 
by a velocity dispersion due to unresolved structures in the resolution element \citep{feldman_1983}, or 
$-$ as discussed by \citet{depontieu_etal_2014} $-$ by a twisting motion of small-scale flux elements. 

The k$_{\rm{2}}$ width of the \ion{Mg}{ii}\,k line in LBs is larger than umbra but comparable to quiet
Sun (Fig.~\ref{fig:hist}, panel \emph{I}). This is in contrast to the width of the  \ion{O}{iv}\,$140.115$\,nm lines
which shows a similar distribution for the umbra, LB, and quiet Sun, although the LB and quiet Sun samples have a larger tail 
toward larger line widths (panel \emph{L}). 
The situations in the \ion{Si}{iv} and \ion{C}{ii} lines are different: while the LB distribution peaks close
to the quiet Sun sample for the former, it peaks close to the umbra sample for the latter (panels \emph{J,\,{\rm and}\,\,K}). 
Distribution of the width of the \ion{C}{ii} line in quiet Sun and sunspot presented here 
are in agreement with \citet{rathore_etal_2015}. 
In both cases, the line width is the result of fitting a single-Gaussian function. 
That the \ion{C}{ii} line width in LBs is more comparable to umbra is consistent with our
finding that the contrast of LBs in this line are not as large as in the \ion{Mg}{ii}\,k
and \ion{Si}{iv} lines (panels \emph{D,\,F\,{\rm and}\,\,G}). 
As discussed by \citet{avrett_loser_07} and \citet{rathore_carlsson_2015}, the formation of the \ion{C}{ii} line pair and also the
background opacity in the $133$\,nm wavelength range should be treated in nonlocal thermodynamic equilibrium.
Three-dimensional effects produce a \ion{C}{ii} line profile that is significantly influenced 
by nearby regions \citep[][their Fig.~10]{rathore_carlsson_2015}. 
The net effect is that we observe LBs that are less distinct
in the \ion{C}{ii} line than they are in the \ion{Mg}{ii}\,k and \ion{Si}{iv} lines.

We found evidence for a spatial correlation between the logarithm of the intensity and the line width
in LBs in all of the lines in this study (Fig.~\ref{fig:intensity_width}). 
Our finding is in agreement with \citet{dere_mason_1993} who found significant correlation between the \ion{Si}{iv} 
intensity and line width in quiet Sun. \citet{peter_2000} reported a similar correlation in the networks. 
A comparable relation between the intensity and line width was 
obtained by \citet{lemaire_etal_1999} for coronal holes and for quiet Sun. 
\citet{akiyama_etal_2005} also found a line width$-$intensity spatial correlation for a set of TR spectral lines in quiet Sun.
They did not find a clear correlation for the \ion{Si}{iv} line in quiet Sun while there is a correlation 
in our LB sample.

In our statistical ensemble of LBs, the line width of the simultaneously observed
FUV lines show a spatial correlation (Fig.~\ref{fig:ww}). The relation between
the line width of the \ion{C}{ii}, \ion{Si}{iv}, and \ion{O}{iv} lines indicates that there is a
correlation between plasma in this range of temperatures ($3\times10^4$ to $1.5\times10^5$K) in a LB atmosphere.
This is not fully consistent with the idea of a one-dimensional stratification in which 
different temperatures occupy different heights, unless one assumes that there is a coherent structure 
which extends over a few Mm. 
Our results do not support a stationary bright wall; unlike the \ion{Mg}{ii}\,k line, the UV continuum 
$-$ and to a minor extent weak FUV lines $-$ do not show the LB pattern in all cases.
If there were a bright wall (hotter than the nearby umbra) from the photosphere to the TR,
the UV continuum would show the LBs in all cases. 
Supporting a solid bright wall is further complicated considering the vast dynamics of LBs in the
photosphere and chromosphere. 
Furthermore, there is an ample amount of fine structure in the chromosphere and TR of LBs, 
which has a small scale down to the resolution element. 
We interpret the correlations as a signature of a multi-component TR. 
The fine structure in LBs and indications of a multi-component atmosphere
support an inhomogeneous TR discussed by \citet{peter_2001}.

\subsection{Doppler shifts}
Light bridges in the TR show both redshift and blueshift, although the redshift is dominant.
Only in the strong photospheric line of the \ion{Fe}{i}\,$283.2435$\,nm and the
chromospheric lines of \ion{Mg}{ii}\,k and \ion{O}{i}\,$135.560$\,nm, 
are the blue- and red-shifts comparable (Fig.~\ref{fig:hist}, panels \emph{M-N}). 
The range of the measured velocities significantly increases with the formation temperature of the FUV lines. 
The median redshift in the \ion{Si}{iv} line in the quiet Sun sample ($7.3$\,km\,s$^{-1}$) is
larger than the corresponding value in the LB and umbra samples ($4.4$ and $2.4$\,km\,s$^{-1}$, respectively).
The average quiet Sun redshift in the sample is slightly larger than
the typical average quiet Sun velocities reported for this line \citep{peter_judge_1999}, 
but this can be due to selection effects as the quiet Sun sample is not very quiet.  
The quiet Sun sample is often in the moat around sunspots.   
The \ion{C}{ii} velocity distribution in the quiet Sun sample (panels \emph{O}) has a median value at
$2.3$\,km\,s$^{-1}$, while the corresponding velocity for the LB / umbra sample is  $2$\,/\,$1.8$\,km\,s$^{-1}$. 

The Doppler shifts of the \ion{Mg}{ii}\,k, \ion{C}{ii}, and \ion{Si}{iv} lines show a positive correlation while
the \ion{O}{i} line-core velocity does not show a clear trend with other lines (Fig.~\ref{fig:velo}).
As mentioned earlier, the  \ion{O}{i} line is weaker than
the \ion{C}{ii} and \ion{Si}{iv} lines and hence is more prone to noise. 
The correlation between the Doppler shifts is in accordance with the correlations between the line widths and the line intensity
discussed above. 
The coupling between the cool and hot plasma is essential in the heating processes. 
In the cool plasma there are many neutrals, while the bulk of the TR consists of ionized atoms. 
Recent numerical models hint at the importance of the ambipolar diffusion in the plasma heating
in  the chromosphere where hydrogen and helium are partially ionized \citep{khomenko_etal_2014, sykora_etal_2015}. 
Altogether, we found evidence for a physical state in LBs where cool and hot plasma
mutually influence each other and their spatial variation shows some degree of coherence. 
We conclude that LBs consists of multi-component (multi-thermal) atmospheres.

\subsection{Energy balance}
IRIS observations of LBs have revealed enhanced emission compared to
nearby umbra from the photosphere to the chromosphere and TR.} 
Although the bulk of LBs are brighter or much brighter than nearby umbra in all lines,
there are some LBs that are not significantly brighter in one or more lines,
in particular in the weak \ion{O}{i} and the UV continuum. In other words, there is an intrinsic diversity 
in the contrast of LBs in the FUV emission lines, while they are always 
distinctly brighter in the \ion{Mg}{ii}\,k line. This perhaps indicates that 
different energy sources govern the cool and hot lines (see below). 
This reminds us of the variation in the continuum contrast of LBs in the photosphere as discussed by
\cite{rolf_etal_2016}. Further investigations are required to find out how far the amount of energy deposited
in the chromosphere and TR depends on the photospheric structure and dynamics. 
Photospheric observations provide evidence for overturning convection in LBs  \citep{rimmele_08, roupe_luis_etal_2010}.
A convectively driven LB is also supported by the numerical simulations of \cite{cheung_etal10} and \citet{rempel_2011}.
These simulations do not extend beyond the temperature minimum and so cannot be directly compared with our results. 
Even if LBs show a close resemblance to quiet Sun in the photosphere, our results prove that their properties
notably deviate from quiet Sun in the chromosphere and TR.
As a result, extra heating mechanisms are required to explain their excess emission 
in addition to those operating in quiet Sun.

The intensity of spectral lines and continuums in LBs compared to quiet Sun varies in the atmosphere.
Although LBs are usually not as bright as granulation in the photosphere, 
they are significantly brighter in the \ion{Mg}{ii}\,k line and at the same time only slightly brighter in the \ion{O}{i} line. 
The \ion{O}{i} line forms at a height of $1.0-1.5$\,Mm \citep{lin_carlsson_2015} 
while the magnesium line has contributions from layers up to several Mm above the solar surface. 
In other words, the contrast of LBs compared to quiet Sun increases 
from photospheric continuum up to a height of about $2.0-2.5$\,Mm 
where the k$_{\rm{3}}$ forms \citep{leenaarts_etal_2013b}. 
The contrast does not further increase upward for the \ion{C}{ii} 
and \ion{Si}{iv} lines and at the same time there is little 
difference between the LB and quiet Sun in the \ion{O}{iv} line (Fig.~\ref{fig:hist}). 
If the energy flow were primarily from the corona 
toward the chromosphere as suggested by the downflow in the TR lines, 
one would expect to find a larger relative intensity in the TR than in the chromosphere. 
Our findings support an energy deposit which peaks in the middle/upper chromosphere.

Oscillations are prevalent in the solar atmosphere. 
Umbral flashes are oscillations observed in the chromosphere of sunspot umbra  \citep{socas_etal_2000}. 
Maps of the chromospheric intensities do not show the umbral flashes in LBs.  
We used IRIS time series to estimate importance of the mechanical heating.  
To this end, we performed a Fourier analysis of a $2$h fixed-slit time series for entries $15$ and $16$ in Table~\ref{tab:sample}. 
A comparison of the Fourier power in the LB and umbra reveals that while umbral flashes have a 
typical frequency higher than $5$\,mHz, the power in LBs peaks 
at frequencies lower than $5$\,mHz both in the intensity and velocity of 
the \ion{Mg}{ii}\,k line. This agrees with the fact that we do not find signs of umbral flashes in LBs. 
Similar results were reported by \citet{zang_etal_2017}, who observed different oscillation frequencies in LBs and umbra.  
\citet{yurchyshyn_etal_2015} found strong power concentration associated with a LB using a $100$ min IRIS time series. 
These findings are comparable to those of \citet{sobotka_etal_2013} who measured the acoustic power of LBs in the photosphere. 
In summary, the oscillation period of LBs observed in the \ion{Mg}{ii}\,k line is between three and five minutes. 
There is also a significant power concentration observed associated with LBs. 
Therefore, dissipation of the mechanical energy contributes to the chromospheric energy balance of LBs.

It is not clear whether the spectral lines discussed here are supplied by a common heating mechanism or 
experience diverse heating regimes.
The strong dependence of the longitudinal thermal conductivity on the temperature \citep[$T^{5/2}$,][]{stix_book} implies 
that the thermal conduction is not efficient for a generic chromospheric temperature of $10^4$\,K. 
The situation changes toward the upper chromosphere where the temperature and the temperature gradient increase substantially. 
The conducted heat flux is concentrated in channels of strong magnetic field strength.
As seen in Fig.~\ref{fig:mask}, the width of the LBs in the TR is comparable to the one in the photosphere, although 
the pressure and density between the two layers are different by several orders of magnitude. 
In other words, LBs do not exponentially expand with height as they are surrounded by strong magnetic field of umbra. 
Furthermore, we do not observe peripheral intensity enhancements in LB maps of the TR lines:
the heating should happen through the volume of the LB and not over its boundary.  
It is feasible to perceive that an efficient thermal conduction in the \ion{Si}{iv} and \ion{O}{iv} lines
contributes to the heating of LBs. An exponentially increasing density with depth along with an inefficient thermal conductivity 
at low temperatures prevents the thermal conduction from playing an important role in the low chromosphere.

In quiet Sun, the upper chromosphere and TR lines form above the $\beta=1$ surface, i.e., in a highly
conductive media in which the magnetic pressure dominates the gas pressure 
\citep[the plasma-$\beta$ is the ratio of the thermal pressure to the magnetic pressure,][]{aschwanden_2004}. 
In sunspot umbra, the surface of equipartition between magnetic and thermal energy density
is located at a lower geometrical height than the quiet Sun. 
\citet{metcalf_etal_1995} found that in a sunspot, the magnetic field becomes force-free in the low chromosphere.  
In contrast, \citet{socas_2005a} observed a sunspot with a spectropolarimeter
in the photosphere and chromosphere and found that the magnetic field was not force-free up to $\approx\!2$Mm.
A LB perhaps further complicates the magnetic field structure and the field deviates from a potential configuration. 
We tried to roughly estimate the plasma-$\beta$ in LBs assuming a field strength of $\approx\!1$\,kG and a
thermodynamic stratification similar to  model-M of \citet{maltby_etal_86}.
The plasma-$\beta$ is smaller than unity for layers at $\log\tau\approx-4$ and above. 
In addition, the convective motion in LBs  
builds up a magnetic free energy \citep{peter_etal_2004}. 
\citet{reza_etal_2012a} discussed evolution of a forming sunspot and its impact on the magnetic field structure of a LB. 
\citet{roubustini_etal_2016} found  magnetic reconnection as the driving force of the small-scale jets in the LBs. 
We conclude that currents should be abundant in LBs. 
At present, there are many reports of significant current in the photosphere \citep{jurcak_etal_2006, felipe_etal_2016} and 
a few in the chromosphere \citep{tritschler_etal_08}. 
In summary, the Joule heating contributes to the emission enhancements in the LBs, which 
is consistent with our finding that the median intensity ratio
of the \ion{C}{i}\,$135.58$\,nm and \ion{O}{i}\,$135.56$\,nm lines 
is about $0.5$ in LBs compared to $\approx\!0.3$ in quiet Sun.
The ratio was interpreted as a tracer of flaring activity by \citet{cheng_etal_1980}. 
To further quantify the importance of the magnetic, mechanical, or thermal conduction energy, 
the field strength and oscillations have to be measured simultaneously. 
In particular, high-resolution ground observations help to distinguish
if segmented or filamentary LBs harbor different heating mechanisms.

\section{Conclusion}\label{sec:conclusion}
We studied a sample of light bridges (LBs) observed with IRIS to address their presence and
properties in the chromosphere and transition region. 
The strength of the emission lines varies among LBs and also depends on the selected spectral line.
Light bridges have a clear signature in the  Mg\,{\sc ii}\,k and \ion{Si}{iv} lines. 
The correlations between the line widths and the intensities of spectral lines originating from 
plasma at different temperatures implies an inhomogeneous and multi-thermal state   
that deviates from a typical quiet Sun and umbra atmosphere. This is further fortified
by strong correlations between the intensity of different lines. 
Considering the small-scale structure of LBs, 
the spatial distribution, and the diversity in the emission enhancements, 
light bridges are  multi-thermal, dynamic, and  coherent structures in the
solar atmosphere from the photosphere to the transition region. 
The observed emission enhancements in the chromosphere and transition region compiles evidence against
the cusp-shape magnetic structure of LBs. 
The variation in the emission enhancements in different layers indicates that 
different heating mechanisms operate in the atmosphere.

\begin{acknowledgements}
IRIS is a NASA small explorer mission developed and operated by LMSAL with mission operations executed at NASA Ames 
Research center and major contributions to downlink communications funded by the Norwegian Space Center (NSC, Norway) 
through an ESA PRODEX contract. The author gratefully acknowledges support from the Spanish Ministry of Economy
and Competitivity through project AYA2014-60476-P (Solar Magnetometry in the Era of Large Solar Telescopes). 
Part of this work was supported by the DFG grant RE\,3282/1-2. 
This work has made use of the VALD database, operated at Uppsala University, the Institute of Astronomy RAS in Moscow, 
and the University of Vienna. We would like to thank the referee for the constructive suggestions.
\end{acknowledgements}

\bibliography{rezabib_nn}

\end{document}